\documentclass[12pt]{article}

\usepackage{sbc-template}
\usepackage{graphicx,url}
\usepackage[utf8]{inputenc}
\usepackage[T1]{fontenc}
\usepackage[brazil]{babel}
\usepackage{hyperref} 
     
\title{Questões Epistemológicas em \\Mineração de Dados Educacionais}

\author{
	Esdras L. Bispo Jr. 
}

\address{
	Instituto de Ciências Exatas e Tecnológicas (ICET) \\Universidade Federal de Goiás (UFG)\\
	\nextinstitute
	Jataí ACM SIGCSE {\it Chapter}\\
	{\it Association for Computing Machinery} (ACM)
  \email{bispojr@ufg.br}
}

\begin{document} 

\maketitle

\begin{abstract}
  Educational Data Mining (EDM) shows interesting scientific results lately. However, little has been discussed about philosophical questions regarding the type of knowledge produced in this area. This paper aims to present two epistemological issues in EDM: (i) a question of ontological nature about the content of the knowledge obtained; and (ii) a question of deontological nature, about the guidelines and principles adopted by the researcher in education, to the detriment of the results of his own research. In the end, some considerations and guidelines are outlined as a result of the discussion of the issues raised.
\end{abstract}
     
\begin{resumo} 
  Resultados cientificamente interessantes vêm sendo apresentados em mineração de dados educacionais (MDE). Entretanto, pouco se tem discutido sobre questões filosóficas em relação ao tipo de conhecimento produzido nesta área. Este trabalho tem como propósito apresentar duas questões epistemológicas em MDE: (i) uma questão de natureza ontológica sobre o conteúdo do conhecimento obtido; e (ii) uma questão de natureza deontológica, sobre as pautas e os princípios adotados pelo pesquisador na educação, em detrimento do resultados de sua própria pesquisa. Ao final, algumas considerações e diretrizes são delineadas como resultado da discussão das questões levantadas.
\end{resumo}

\section{Introdução}

A Mineração de Dados Educacionais (MDE) é uma área emergente de pesquisa interdisciplinar que lida com o desenvolvimento de métodos para explorar dados originários em um contexto educacional \cite{romero:2010}. Alguns resultados da área vão em direção à caracterização da (i) evasão \cite{manhaes:2011, nandeshwar:2011}, (ii) motivação \cite{winne:2013, santos:2015} e (iii) desempenho de estudantes \cite{gottardo:2012, sullare:2016}.

A MDE utiliza abordagens computacionais para analisar dados educacionais com o objetivo de estudar questões educacionais. Entretanto, pouco se tem discutido sobre questões filosóficas em relação ao tipo de conhecimento produzido nesta área. Este tipo de problematização não é devidamente tratado nos trabalhos de revisão sistemática publicados nos últimos anos \cite{coelho:2017, li:2015, papamitsiou:2014, pena:2014, romero:2010, sergis:2017, vahdat:2015}. Existe uma menção bastante breve em \cite{clow:2013} sobre a fragilidade da epistemologia do {\it learning analytics}.

Entretanto, questões epistemológicas envolvendo algumas asserções científicas em relação ao {\it Big Data}, por exemplo, existem na literatura \cite{leonelli:2014, boyd:2012}. Nas Ciências Biológicas, é comum o uso de grande massa de dados para realização de pesquisas empíricas (e.g. Bases do NCBI\footnote{{\it National Center for Biotechnology Information} (NCBI): \url{https://www.ncbi.nlm.nih.gov}.}). Questionamentos sobre o tipo de conhecimento gerado pelo {\it Big Data}, e sobre a sua relevância científica, têm sido apontados e discutidos nestes trabalhos. 

Este trabalho tem como propósito apresentar duas questões epistemológicas em MDE: (i) uma questão de natureza ontológica sobre o conteúdo do conhecimento obtido; e (ii) uma questão de natureza deontológica, sobre as pautas e os princípios adotados pelo pesquisador da informática na educação, em detrimento do resultados de sua própria pesquisa.

O restante do trabalho está dividido como se segue. A Seção \ref{sec:md-e-ed} descreve os principais conceitos da MDE. A Seção \ref{sec:epis} apresenta a epistemologia e suas perguntas norteadoras. A Seção \ref{sec:quest} estrutura duas questões epistemológicas em MDE. E, por fim, na Seção \ref{sec:cons}, considerações finais e diretrizes são delineadas em relação ao que é discutido no corpo deste trabalho.

\section{Mineração de Dados e Educação} \label{sec:md-e-ed}

Uma das principais justificativas para o surgimento da mineração de dados (MD) é a disponibilidade. A informatização dos dados possibilitou um acesso a uma vasta quantidade de dados que antes não era disponível. As organizações, através de seus gestores e/ou funcionários, têm acesso a este grande volume de dados a um baixo custo \cite{mitra:2002}.

Um dos pressupostos da MD é que existem informações implícitas dentro desta grande massa de dados \cite{fayyad:1996}. Um processo diretamente associado a MD é a descoberta do conhecimento (DC)\footnote{Da expressão, em inglês, {\it Knowledge Discovery} (KD)}. A definição de DC é a extração de informação implícita, previamente desconhecida e potencialmente útil, a partir de dados \cite{bramer:2007}. A extração destas informações é realizada através de técnicas específicas, como por exemplo, aquelas relacionadas ao reconhecimento de padrões \cite{pal:2004}.

A MD tem aplicações nas mais variadas esferas de conhecimento. Dentre elas, pode-se citar, na esfera da ciência, as análises de imagens de satélites e de compostos orgânicos, sistemas de recomendação em diagnósticos médicos, e análise de dados do processo de aprendizagem . Na esfera do mercado, temos, como exemplos, a detecção automática de fraude em cartões de créditos, análise de sentimentos (de satisfação) de usuários, previsão de séries temporais financeiras (e.g. bolsa de valores), e identificação de público-alvo para um determinado setor.

Um contexto mais específico é a Educação. A área de Mineração de Dados Educacionais (MDE) surge possivelmente com a publicação do trabalho de \cite{anjewierden:2007}. Este trabalho utilizou MD para identificar perfis de estudantes. Desde então, uma gama de pesquisas vem sendo realizada a partir deste recorte.

Os pesquisadores em MDE utilizam uma variedade de fontes de dados para realizar as suas pesquisas \cite[p.~xi]{romero:2010b}. Estes dados são oriundos de tutores inteligentes, sistemas educacionais baseados em computadores, fóruns de discussões online, diários de classes eletrônicos, registros acadêmicos digitais de estudantes, entre outros. Este grande volume de dados tem proporcionado novos caminhos de se estudar o processo de aprendizagem.

A MDE tem contribuido em diversas direções. Podemos citar alguns deles, tais como (i) a visualização de grande volume de informações, (ii) a análise estatística de dados provenientes de ambientes virtuais de aprendizagem (AVA); e a produção de (iii) classificações, (iv) agrupamentos, e (v) regras de associação em repositórios de dados educacionais \cite[p.~3-4]{romero:2010b}.

\section{Conceitos em Epistemologia} \label{sec:epis}

A Epistemologia é um ramo da filosofia conhecido como Teoria do Conhecimento \cite{rescher:2012}. Ela é o estudo do conhecimento e das crenças justificadas \cite{steup:2017}. Seu surgimento acontece na Grécia antiga, juntamente com o surgimento da Filosofia \cite{gerson:2009}.
 
Um conjunto de perguntas são de interesse da epistemologia. Podemos citar algumas delas, tais como: (i) ``Quais são as condições necessárias e suficientes do conhecimento?''; (ii) ``Quais são as suas fontes?''; (iii) ``Qual é a sua estrutura?”; (iv) “Quais são os seus limites?'' \cite{steup:2017}. A pergunta ``o que se pode conhecer?'' também é um tópico de interesse da epistemologia.

A Epistemologia fornece um ferramental necessário para uma autocrítica da pesquisa científica. Ao se questionar sobre as fontes do conhecimento e seus limites, a Ciência tem a oportunidade de se revisitar e estruturar adequadamente a natureza do conhecimento produzido por ela. A Epistemologia promove, como uma de suas áreas, o que é conhecida como Filosofia da Ciência, possibilitando um processo de refinamento e evolução do ensino das ciências e dos paradigmas científicos que surgiram (e possivelmente surgirão) na história \cite{burbules:1991}.

É interessante ressaltar que perguntas semelhantes às da Epistemologia têm sido feitas pela Ciência da Computação. Pode-se dar como exemplos as perguntas realizadas pela Teoria dos Autômatos: ``Quais são as capacidades e limitações fundamentais dos computadores?'' \cite{sipser:2012}. Todavia, ao invés de desejar compreender os limites da computabilidade, a Epistemologia visa conhecer os limites do conhecimento.

Para a MDE, a Epistemologia pode fornecer um arcabouço importante na compreensão dos resultados por ela gerados. Existe um incentivo cada vez maior do uso da tecnologia em ambientes educacionais. E, diante disso, é necessário compreender quando, como e em que condições a inclusão da MDE pode fornecer uma contribuição científica relevante para a área.

Serão apresentadas e discutidas a seguir algumas relações entre a epistemologia e a cosmovisão pedagógica (Seção \ref{subsec:epis-cosm}) e entre a epistemologia e a deontologia (Seção \ref{subsec:epis-deon}). 

\subsection{Epistemologia e Cosmovisão Pedagógica} \label{subsec:epis-cosm}

Algumas questões epistemológicas são desenvolvidas com o propósito de se estabelecer uma cosmovisão pedagógica. É possível identificar este esforço, por exemplo, no trabalho de \cite{morin:2014}. Sobre os sete saberes necessários à educação do futuro, Morin aborda questões de ordem epistemológica em pelo menos três deste saberes.

O primeiro saber necessário, apontado por Morin, está associado às cegueiras do conhecimento. O autor afirma que a função da educação é ``mostrar que não há conhecimento que não esteja, em algum grau, ameaçado pelo erro e pela ilusão''. Faz parte da educação ensinar os limites do próprio conhecimento, apontando os seus próprios pontos fracos, suas próprias fragilidades.

Um outro saber necessário, listado por Morin, são os princípios do conhecimento pertinente. Afirma o autor que a educação necessita ``situar tudo no contexto e no complexo planetário''. O pensamento complexo \cite{morin:2007} é um dos elementos que norteia a proposta de educação de Morin. É necessário que o conhecimento seja interligado e articulado com o todo, evitando uma educação fragmentada, descontextualizada do mundo.

E, por fim, mais um outro saber necessário, identificado pelo autor, é o ensinar a compreensão. Morin aponta quais seriam as condições necessárias para o conhecimento em termos humanísticos. O ensino da compreensão em uma era global, interconectada por redes e pela mídia, é ``condição e garantia da solidariedade intelectual e moral da humanidade''. Neste ponto, percebe-se a ligação estreita que existe entre aspectos epistemológicos e deontológicos \cite{muller:2007}.

\subsection{Epistemologia e Deontologia} \label{subsec:epis-deon}

Conforme identificado anteriormente, existe uma proximidade entre questões epistemológicas e deontológicas. A deontologia é um dos tipos de teorias normativas relacionadas com as escolhas que são moralmente exigidas, proibidas ou permitidas \cite{alexander:2016}. Geralmente a deontologia é tratada em conjunto com a área da Ética e está preocupada com questões associadas com o dever.

A pergunta epistemológica ``o que se pode conhecer?'' pode ter uma percepção deontológica sob várias circunstâncias. Uma delas está associada com as limitações relacionadas ao dever. Por exemplo, admita que uma pessoa possa ter acesso ao diário de anotações pessoais de um amigo próximo. O acesso àquela informação pode até existir de fato. Entretanto, esta pessoa pode recusar-se a lê-lo por questões de respeito ao amigo, motivadas talvez pela existência de um protocolo social.

Logo, as limitações do acesso a esta informação pode não ser de ordem física ou cognitiva. Elas podem ser limitações de ordem normativa. E esta norma pode ser uma imposição de natureza intrínseca ou extrínseca ao indivíduo. Isto é, pode existir coação (ou compreensão) externa para o cumprimento deste dever; ou pode existir compreensão (ou coação) interna de que aquele limite deve ser respeitado.

Em Educação, é bastante comum vermos um professor adotar uma determinada pedagogia de trabalho. Esta pedagogia guia o professor em sua prática profissional através de princípios que o levam como enxergar tanto seu espaço de trabalho como todos os envolvidos nele. Algumas escolhas realizadas na esfera de sua prática são tomadas em consonância com esta ``cosmovisão educacional'' adotada por este profissional.

Para esta pedagogia de trabalho, em \cite{ernest:2012} e \cite{ben:2001}, o termo ``paradigma educacional'' é utilizado. Para os autores, um paradigma educacional é composto de quatro elementos:
\begin{enumerate}
	\item uma ontologia que é uma teoria da existência;
	\item um epistemologia que é uma teoria do conhecimento, tanto do conhecimento específico de um indivíduo tanto conhecimento humano compartilhado;
	\item uma metodologia para aquisição e validação do conhecimento; e
	\item uma pedagogia que é uma teoria de ensino, que significa facilitar o aprendizado de acordo com a epistemologia.
\end{enumerate}

Desta forma, é de se esperar naturalmente que o exercício profissional do professor dependa diretamente do paradigma educacional adotado por ele. Uma vez que ele adote este paradigma, os seus princípios e pressupostos escolhidos servirão para ele como norteadores deontológicos de sua prática docente. Assim, questões epistemológicas têm uma relação de proximidade com questões deontológicas na Educação. 

\section{Questões Epistemológicas em MDE} \label{sec:quest}

Serão apresentadas e discutidas a seguir duas questões epistemológicas em MDE. A primeira diz respeito à natureza ontológica do conteúdo do conhecimento obtido na área (Seção \ref{subsec:q-onto}). E a segunda diz respeito à natureza deontológica que diz respeito às pautas e aos princípios adotados pelo pesquisador na Educação (Seção \ref{subsec:q-deon}).

\subsection{Questão de Natureza Ontológica} \label{subsec:q-onto}

Um aspecto importante a salientar em MD é a natureza ontológica do conhecimento gerado. Em MD, o conhecimento é gerado a partir do fornecimento de dados a um algoritmo. A partir destes dados, padrões são obtidos por meio de técnicas específicas de classificação. 

Um aspecto interessante é a natureza destas técnicas de classificação de padrões. Todas elas são construídas sobre uma perspectiva indutiva do conhecimento. São a partir daqueles dados que os padrões são justificados e determinados. Estes padrões comumente são utilizados para fins preditivos em um processamento de dados no futuro.

A questão que emerge deste fato é sobre como estão sendo utilizados os resultados gerados em MDE. Temos, ao menos, duas fontes de problemas relacionados à natureza deste conhecimento gerado, ambas oriundas da falta de representatividade estatística dos dados, os quais apresentaremos a seguir.

O primeiro problema diz respeito à dimensão espacial dos dados. Os padrões gerados refletem características pertinentes àqueles dados extraídos. E não necessariamente refletem às características existentes na população de elementos como um todo. É necessário conhecer o poder que o conhecimento de natureza indutiva proporciona. E também é necessário compreender as limitações a ele associadas. É possível que, embora existam muitos dados disponíveis para serem utilizados em algum algoritmo de MDE, estes dados podem ser pouco representativos para fornecer embasamento, produzindo um padrão que não represente adequadamente o contexto específico ao qual se queira observar.

Um cenário ilustrativo seria o de utilizar todos os dados gerados em um AVA para uma MDE. Admita que a partir deste dados, deseja-se prever determinado comportamento dos usuários, observando os dados gerados pelos acessos a este AVA. Entretanto, imagine que a maioria destes acessos foram realizados por estudantes de ensino fundamental, em relação aos estudantes de ensino médio. Os padrões gerados por um algoritmo em MDE será, neste caso, fortemente enviesado pela quantidade de dados superior de estudantes do nível fundamental, em detrimento de estudantes do nível médio. É necessário haver uma leitura crítica dos resultados gerados pelo algoritmo ou uma seleção prévia dos dados que serão utilizados na etapa de reconhecimento de padrões.

O segundo problema diz respeito à dimensão temporal dos dados. Embora não haja uma falta de representatividade estatística aparente dos dados, é possível que o momento em que foi realizada a coleta dos dados afete significativamente o poder de previsão do padrão gerado. Uma generalização de um padrão pode ser apropriada para a previsão de pontos de séries temporais próximos do momento de extração destes dados. Entretanto, a distância temporal entre a coleta do dado e o uso do padrão gerado em MDE pode destoar de tal forma que aqueles dados podem perder naturalmente a sua capacidade preditiva pelo simples fato de estar suficiente distante do contexto. 

Um possível cenário seria uma universidade que tenha acesso a um grande volume de dados sobre os seus estudantes desde a sua fundação. Ao informatizar todos estes dados, deseja-se construir o perfil do estudante ingressante em determinado curso através de algoritmos de MD. Imagine que o horizonte de dados coletados foi de um período de 50 anos. É bem possível que os motivos que levaram os estudantes a escolher determinado curso variam significativamente dentro deste horizonte. Extrair um perfil de estudante considerando todo este universo, desconsiderando as diversas particularidades temporais existentes fatalmente levará o gestor educacional a uma compreensão equivocada da informação que de fato ele necessita obter. 

O intrigante é que a maior parte destes problemas já foram devidamente tratados por diversos autores, como por exemplo, em relatos de falácias indutivas em Lógica (Salmon, 2010). O problema crucial em relação à MDE são os pressupostos implícitos que muitas pessoas carregam consigo, tanto em relação à tecnologia quanto em relação ao fascínio existente com o grande volume de dados \cite{lazer:2014}. A admissão simples de que a imensa quantidade de dados, em conjunto com a tecnologia, faz emergir naturalmente o conhecimento de que os gestores educacionais realmente necessitam é uma das principais responsáveis por conduzir alguns pesquisadores a estes equívocos. 

Além destas questões, ainda existem outras questões de ordem ontológica. O processo de ``descoberta do conhecimento'' também é um aspecto, às vezes, mal compreendido por pesquisadores em Educação. Há uma discussão específica sobre a natureza do conhecimento e da sua relação intrínseca a seres humanos (e não a máquinas). Para \cite{setzer:2005}, o conhecimento está intimamente ligado à informação e à semântica, sendo uma atividade passível de ser executada exclusivamente por seres humanos. Nesta direção, o que as máquinas podem fazer apenas é o processamento de dados (e não de informação). E, como consequência natural, as máquinas não poderiam realizar a ``descoberta do conhecimento''. 

Mesmo ainda sob esta ótica, é possível utilizar a expressão ``descoberta do conhecimento''. Basta enxergarmos a MDE sob a ótica de um sistema de informação, de forma que a tecnologia (a MDE) é uma parte de sua composição \cite{laudon:2016}. Todas as pessoas envolvidas também seriam parte da MDE, juntamente com todas as inter-relações entre máquinas, pessoas e organizações. 

\subsection{Questão de Natureza Deontológica} \label{subsec:q-deon}

Um outro aspecto importante a salientar em MDE é a natureza deontológica do conhecimento gerado. Normalmente, professores, estudantes, gestores e demais atores envolvidos no contexto educacional utilizam este conhecimento como insumo no processo de tomada de decisões. A MDE situa-se normalmente, dentro deste processo, como uma provedora de informações diferenciadas, sendo, às vezes, estas informações as mais decisivas diante das demais.

Nesta direção, é necessário perceber os possíveis conflitos em termos deontológicos que o conhecimento descoberto pela MDE pode ocasionar. As informações sugeridas podem antagonizar com princípios morais e/ou deveres e protocolos socialmente aceitos dentro de determinadas instituições. Algumas sugestões oriundas da MDE podem até colidir com valores culturais construídos e consolidados por uma comunidade.

Em \cite{caliskan:2017}, apresenta-se os riscos existentes da descoberta de conhecimento automática a partir de dados produzidos por humanos em linguagem natural. Em alguns cenários, os autores indicaram que os algoritmos podem reproduzir preconceitos e estereótipos existentes na sociedade, como o racismo e o machismo. Por exemplo, os nomes de origem africana foram mais associados a palavras negativas do que os nomes de origem europeia.

Desta forma, é necessário estar atento aos tipos de conhecimento que estão sendo descobertos ao se utilizar a MDE. A utilização da MDE pode, se utilizada sem o devido cuidado, contrariar os valores expressos e defendidos em um projeto político pedagógico, por exemplo. Os padrões obtidos costumam refletir mais fielmente à realidade, tal como ela é, do que aos valores desejados e buscados por uma determinada comunidade escolar. Saber ler e interpretar o resultado apresentado por tais algoritmos é essencial para a exploração de todo o potencial que a MDE tem a oferecer.

Outros aspectos em relação à transparência e à governança destes sistemas que utilizam MD têm sido discutidos recentemente. A necessidade de se criar um ``contrato social algorítmico'' \cite{rahwan:2018} está surgindo como uma necessidade de se regular estes sistemas que envolvem os mais diversos valores com diversos atores da sociedade. A MDE, sensível a estes esforços já existentes, necessitará em breve percorrer mais solidamente também nesta direção.

\section{Considerações Finais} \label{sec:cons}

Este trabalho teve como propósito apresentar duas questões epistemológicas em MDE: (i) uma questão de natureza ontológica sobre o conteúdo do conhecimento obtido; e (ii) uma questão de natureza deontológica, sobre as pautas e os princípios adotados pelo pesquisador da informática na educação, em detrimento dos resultados de sua própria pesquisa. Algumas armadilhas sobre o uso da MDE foram apresentadas, pontuando os riscos existentes em cada uma das dimensões pontuadas.

Diante disso, é necessário reafirmar a importância e o potencial que a MDE tem para contribuir com a proposição de artefatos mediadores na Educação. As novas tecnologias de comunicação permitiram novas mediações no cenário educacional como um todo \cite[p.~45]{kenski:2012}. Desprezar a sua importância é negar toda a riqueza de possibilidades que as novas tecnologias podem proporcionar para a construção de uma realidade educacional coerente com os valores defendidos por nossa sociedade hoje e pela sociedade que nós queremos que um dia venha a ser.

Entretanto é necessário utilizar estas novas tecnologias, incluindo a MD, de forma crítica e consciente de seus poderes e limitações. O espírito crítico e moderadamente cético difundido na ciência deve ser defendido como valor essencial para que constantemente revisitemos os nossos achados de pesquisa e reelaboremos o nosso conhecimento, as nossas crenças e os nosso valores de forma coerente. A falácia de que a tecnologia surge como uma espécie de “Messias” para a resolução de todos os problemas existentes da humanidade, incluindo os da Educação, pode levar sutilmente pesquisadores a conclusões infundadas. Estas conclusões podem prestar um desserviço à MDE, principalmente quando o que se é prometido como resultado não consegue ser entregue de fato.

A diretriz principal diante deste cenário é estabelecer um exercício contínuo de reconciliação das descobertas encontradas em MDE com toda a rica discussão já apresentada pela Filosofia da Ciência sobre a natureza do conhecimento e o seu alcance. Este exercício contínuo reforça os avanços alcançados pela área, garantindo a credibilidade das descobertas diante de todos os envolvidos na Educação.

\bibliographystyle{sbc}
\bibliography{sbc-template}

\end{document}